

%
%
%
\def\unredoffs{} \def\redoffs{\voffset=-.31truein\hoffset=-.59truein}
\def\speclscape{\special{ps: landscape}}
%
%
%
%
\newbox\leftpage \newdimen\fullhsize \newdimen\hstitle \newdimen\hsbody
\tolerance=1000\hfuzz=2pt
\catcode`\@=11 
\def\bigans{b }
\def\answ{b }
\ifx\answ\bigans\message{(This will come out unreduced.}
\magnification=1200\unredoffs\baselineskip=16pt plus 2pt minus 1pt
\hsbody=\hsize \hstitle=\hsize 
\else\message{(This will be reduced.} \let\l@r=L
\magnification=1000\baselineskip=16pt plus 2pt minus 1pt \vsize=7truein
\redoffs \hstitle=8truein\hsbody=4.75truein\fullhsize=10truein\hsize=\hsbody
\output={\ifnum\pageno=0 
  \shipout\vbox{\speclscape{\hsize\fullhsize\makeheadline}
    \hbox to \fullhsize{\hfill\pagebody\hfill}}\advancepageno
  \else
  \almostshipout{\leftline{\vbox{\pagebody\makefootline}}}\advancepageno
  \fi}
\def\almostshipout#1{\if L\l@r \count1=1 \message{[\the\count0.\the\count1]}
      \global\setbox\leftpage=#1 \global\let\l@r=R
 \else \count1=2
  \shipout\vbox{\speclscape{\hsize\fullhsize\makeheadline}
      \hbox to\fullhsize{\box\leftpage\hfil#1}}  \global\let\l@r=L\fi}
\fi
%
\newcount\yearltd\yearltd=\year\advance\yearltd by -1900

\def\Title#1#2{\nopagenumbers\abstractfont\hsize=\hstitle\rightline{#1}%
\vskip 1in\centerline{\titlefont #2}\abstractfont\vskip .5in\pageno=0}
\def\Date#1{\vfill\leftline{#1}\tenpoint\supereject\global\hsize=\hsbody%
\footline={\hss\tenrm\folio\hss}}
%

\def\draftmode{\message{ DRAFTMODE }\def\draftdate{{\rm preliminary draft:
\number\month/\number\day/\number\yearltd\ \ \hourmin}}%
\headline={\hfil\draftdate}\writelabels\baselineskip=20pt plus 2pt minus 2pt
 {\count255=\time\divide\count255 by 60 \xdef\hourmin{\number\count255}
  \multiply\count255 by-60\advance\count255 by\time
  \xdef\hourmin{\hourmin:\ifnum\count255<10 0\fi\the\count255}}}
\def\nolabels{\def\wrlabeL##1{}\def\eqlabeL##1{}\def\reflabeL##1{}}
\def\writelabels{\def\wrlabeL##1{\leavevmode\vadjust{\rlap{\smash%
{\line{{\escapechar=` \hfill\rlap{\sevenrm\hskip.03in\string##1}}}}}}}%
\def\eqlabeL##1{{\escapechar-1\rlap{\sevenrm\hskip.05in\string##1}}}%
\def\reflabeL##1{\noexpand\llap{\noexpand\sevenrm\string\string\string##1}}}
\nolabels
%
\global\newcount\secno \global\secno=0
\global\newcount\meqno \global\meqno=1
\def\newsec#1{\global\advance\secno by1\message{(\the\secno. #1)}
\global\subsecno=0\eqnres@t\noindent{\bf\the\secno. #1}
\writetoca{{\secsym} {#1}}\par\nobreak\medskip\nobreak}
\def\eqnres@t{\xdef\secsym{\the\secno.}\global\meqno=1\bigbreak\bigskip}
\def\sequentialequations{\def\eqnres@t{\bigbreak}}\xdef\secsym{}
\global\newcount\subsecno \global\subsecno=0
\def\subsec#1{\global\advance\subsecno by1\message{(\secsym\the\subsecno.
#1)}
\ifnum\lastpenalty>9000\else\bigbreak\fi
\noindent{\it\secsym\the\subsecno. #1}\writetoca{\string\quad
{\secsym\the\subsecno.} {#1}}\par\nobreak\medskip\nobreak}
\def\appendix#1#2{\global\meqno=1\global\subsecno=0\xdef\secsym{\hbox{#1.}}
\bigbreak\bigskip\noindent{\bf Appendix #1. #2}\message{(#1. #2)}
\writetoca{Appendix {#1.} {#2}}\par\nobreak\medskip\nobreak}
%
%
\def\eqnn#1{\xdef #1{(\secsym\the\meqno)}\writedef{#1\leftbracket#1}%
\global\advance\meqno by1\wrlabeL#1}
\def\eqna#1{\xdef #1##1{\hbox{$(\secsym\the\meqno##1)$}}
\writedef{#1\numbersign1\leftbracket#1{\numbersign1}}%
\global\advance\meqno by1\wrlabeL{#1$\{\}$}}
\def\eqn#1#2{\xdef #1{(\secsym\the\meqno)}\writedef{#1\leftbracket#1}%
\global\advance\meqno by1$$#2\eqno#1\eqlabeL#1$$}
%
\newskip\footskip\footskip14pt plus 1pt minus 1pt 
\def\footnotefont{\ninepoint}\def\f@t#1{\footnotefont #1\@foot}
\def\f@@t{\baselineskip\footskip\bgroup\footnotefont\aftergroup\@foot\let\next}
\setbox\strutbox=\hbox{\vrule height9.5pt depth4.5pt width0pt}
\global\newcount\ftno \global\ftno=0
\def\foot{\global\advance\ftno by1\footnote{$^{\the\ftno}$}}
%
\newwrite\ftfile
\def\footend{\def\foot{\global\advance\ftno by1\chardef\wfile=\ftfile
$^{\the\ftno}$\ifnum\ftno=1\immediate\openout\ftfile=foots.tmp\fi%
\immediate\write\ftfile{\noexpand\smallskip%
\noexpand\item{f\the\ftno:\ }\pctsign}\findarg}%
\def\footatend{\vfill\eject\immediate\closeout\ftfile{\parindent=20pt
\centerline{\bf Footnotes}\nobreak\bigskip\input foots.tmp }}}
\def\footatend{}
%
%
\global\newcount\refno \global\refno=1
\newwrite\rfile
\def\ref{[\the\refno]\nref}
\def\nref#1{\xdef#1{[\the\refno]}\writedef{#1\leftbracket#1}%
\ifnum\refno=1\immediate\openout\rfile=refs.tmp\fi
\global\advance\refno by1\chardef\wfile=\rfile\immediate
\write\rfile{\noexpand\item{#1\ }\reflabeL{#1\hskip.31in}\pctsign}\findarg}
\def\findarg#1#{\begingroup\obeylines\newlinechar=`\^^M\pass@rg}
{\obeylines\gdef\pass@rg#1{\writ@line\relax #1^^M\hbox{}^^M}%
\gdef\writ@line#1^^M{\expandafter\toks0\expandafter{\striprel@x #1}%
\edef\next{\the\toks0}\ifx\next\em@rk\let\next=\endgroup\else\ifx\next\empty%
\else\immediate\write\wfile{\the\toks0}\fi\let\next=\writ@line\fi\next\relax}}
\def\striprel@x#1{} \def\em@rk{\hbox{}}
\def\lref{\begingroup\obeylines\lr@f}
\def\lr@f#1#2{\gdef#1{\ref#1{#2}}\endgroup\unskip}
\def\semi{;\hfil\break}
\def\addref#1{\immediate\write\rfile{\noexpand\item{}#1}} 
\def\startrefs#1{\immediate\openout\rfile=refs.tmp\refno=#1}
\def\xref{\expandafter\xr@f}\def\xr@f[#1]{#1}
\def\refs#1{\count255=1[\r@fs #1{\hbox{}}]}
\def\r@fs#1{\ifx\und@fined#1\message{reflabel \string#1 is undefined.}%
\nref#1{need to supply reference \string#1.}\fi%
\vphantom{\hphantom{#1}}\edef\next{#1}\ifx\next\em@rk\def\next{}%
\else\ifx\next#1\ifodd\count255\relax\xref#1\count255=0\fi%
\else#1\count255=1\fi\let\next=\r@fs\fi\next}
%

%
\newwrite\ffile\global\newcount\figno \global\figno=1
\def\fig{fig.~\the\figno\nfig}
\def\nfig#1{\xdef#1{fig.~\the\figno}%
\writedef{#1\leftbracket fig.\noexpand~\the\figno}%
\ifnum\figno=1\immediate\openout\ffile=figs.tmp\fi\chardef\wfile=\ffile%
\immediate\write\ffile{\noexpand\medskip\noexpand\item{Fig.\ \the\figno. }
\reflabeL{#1\hskip.55in}\pctsign}\global\advance\figno by1\findarg}
\def\xfig{\expandafter\xf@g}\def\xf@g fig.\penalty\@M\ {}
\def\figs#1{figs.~\f@gs #1{\hbox{}}}
\def\f@gs#1{\edef\next{#1}\ifx\next\em@rk\def\next{}\else
\ifx\next#1\xfig #1\else#1\fi\let\next=\f@gs\fi\next}
\newwrite\lfile
{\escapechar-1\xdef\pctsign{\string\%}\xdef\leftbracket{\string\{}
\xdef\rightbracket{\string\}}\xdef\numbersign{\string\#}}

\def\writestop{\def\writestoppt{\immediate\write\lfile{\string\pageno%
\the\pageno\string\startrefs\leftbracket\the\refno\rightbracket%
\string\def\string\secsym\leftbracket\secsym\rightbracket%
\string\secno\the\secno\string\meqno\the\meqno}\immediate\closeout\lfile}}
\def\writestoppt{}\def\writedef#1{}
\def\seclab#1{\xdef #1{\the\secno}\writedef{#1\leftbracket#1}\wrlabeL{#1=#1}}
\def\subseclab#1{\xdef #1{\secsym\the\subsecno}%
\writedef{#1\leftbracket#1}\wrlabeL{#1=#1}}
\newwrite\tfile \def\writetoca#1{}
\def\leaderfill{\leaders\hbox to 1em{\hss.\hss}\hfill}
\def\writetoc{\immediate\openout\tfile=toc.tmp
   \def\writetoca##1{{\edef\next{\write\tfile{\noindent ##1
   \string\leaderfill {\noexpand\number\pageno} \par}}\next}}}

%
\catcode`\@=12 
%
\edef\tfontsize{\ifx\answ\bigans scaled\magstep3\else scaled\magstep4\fi}
\font\titlerm=cmr10 \tfontsize \font\titlerms=cmr7 \tfontsize
\font\titlermss=cmr5 \tfontsize \font\titlei=cmmi10 \tfontsize
\font\titleis=cmmi7 \tfontsize \font\titleiss=cmmi5 \tfontsize
\font\titlesy=cmsy10 \tfontsize \font\titlesys=cmsy7 \tfontsize
\font\titlesyss=cmsy5 \tfontsize \font\titleit=cmti10 \tfontsize
\skewchar\titlei='177 \skewchar\titleis='177 \skewchar\titleiss='177
\skewchar\titlesy='60 \skewchar\titlesys='60 \skewchar\titlesyss='60
\def\titlefont{\def\rm{\fam0\titlerm}
\textfont0=\titlerm \scriptfont0=\titlerms \scriptscriptfont0=\titlermss
\textfont1=\titlei \scriptfont1=\titleis \scriptscriptfont1=\titleiss
\textfont2=\titlesy \scriptfont2=\titlesys \scriptscriptfont2=\titlesyss
\textfont\itfam=\titleit \def\it{\fam\itfam\titleit}\rm}
 \ifx\answ\bigans\else scaled\magstep1\fi
\ifx\answ\bigans\def\abstractfont{\tenpoint}\else
\font\abssl=cmsl10 scaled \magstep1
\font\absrm=cmr10 scaled\magstep1 \font\absrms=cmr7 scaled\magstep1
\font\absrmss=cmr5 scaled\magstep1 \font\absi=cmmi10 scaled\magstep1
\font\absis=cmmi7 scaled\magstep1 \font\absiss=cmmi5 scaled\magstep1
\font\abssy=cmsy10 scaled\magstep1 \font\abssys=cmsy7 scaled\magstep1
\font\abssyss=cmsy5 scaled\magstep1 \font\absbf=cmbx10 scaled\magstep1
\skewchar\absi='177 \skewchar\absis='177 \skewchar\absiss='177
\skewchar\abssy='60 \skewchar\abssys='60 \skewchar\abssyss='60
\def\abstractfont{\def\rm{\fam0\absrm}
\textfont0=\absrm \scriptfont0=\absrms \scriptscriptfont0=\absrmss
\textfont1=\absi \scriptfont1=\absis \scriptscriptfont1=\absiss
\textfont2=\abssy \scriptfont2=\abssys \scriptscriptfont2=\abssyss
\textfont\itfam=\bigit \def\it{\fam\itfam\bigit}\def\footnotefont{\tenpoint}%
\textfont\slfam=\abssl \def\sl{\fam\slfam\abssl}%
\textfont\bffam=\absbf \def\bf{\fam\bffam\absbf}\rm}\fi
\def\tenpoint{\def\rm{\fam0\tenrm}
\textfont0=\tenrm \scriptfont0=\sevenrm \scriptscriptfont0=\fiverm
\textfont1=\teni  \scriptfont1=\seveni  \scriptscriptfont1=\fivei
\textfont2=\tensy \scriptfont2=\sevensy \scriptscriptfont2=\fivesy
\textfont\itfam=\tenit
\def\it{\fam\itfam\tenit}\def\footnotefont{\ninepoint}%
\textfont\bffam=\tenbf \def\bf{\fam\bffam\tenbf}\def\sl{\fam\slfam\tensl}\rm}
\font\ninerm=cmr9 \font\sixrm=cmr6 \font\ninei=cmmi9 \font\sixi=cmmi6
\font\ninesy=cmsy9 \font\sixsy=cmsy6 \font\ninebf=cmbx9
\font\nineit=cmti9 \font\ninesl=cmsl9 \skewchar\ninei='177
\skewchar\sixi='177 \skewchar\ninesy='60 \skewchar\sixsy='60
\def\ninepoint{\def\rm{\fam0\ninerm}
\textfont0=\ninerm \scriptfont0=\sixrm \scriptscriptfont0=\fiverm
\textfont1=\ninei \scriptfont1=\sixi \scriptscriptfont1=\fivei
\textfont2=\ninesy \scriptfont2=\sixsy \scriptscriptfont2=\fivesy
\textfont\itfam=\ninei \def\it{\fam\itfam\nineit}\def\sl{\fam\slfam\ninesl}%
\textfont\bffam=\ninebf \def\bf{\fam\bffam\ninebf}\rm}
%
%

\hyphenation{anom-aly anom-alies coun-ter-term coun-ter-terms}
\def\inv{^{\raise.15ex\hbox{${\scriptscriptstyle -}$}\kern-.05em 1}}

\def\Dsl{\,\raise.15ex\hbox{/}\mkern-13.5mu D} 
\def\dsl{\raise.15ex\hbox{/}\kern-.57em\partial}

\font\bigit=cmti10 scaled \magstep1
\def\lspace{\ifx\answ\bigans{}\else\qquad\fi}
\def\lbspace{\ifx\answ\bigans{}\else\hskip-.2in\fi} 
\def\boxeqn#1{\vcenter{\vbox{\hrule\hbox{\vrule\kern3pt\vbox{\kern3pt
           \hbox{${\displaystyle #1}$}\kern3pt}\kern3pt\vrule}\hrule}}}
\def\mbox#1#2{\vcenter{\hrule \hbox{\vrule height#2in
               \kern#1in \vrule} \hrule}}  
%

\def\e#1{{\rm e}^{^{\textstyle#1}}}

\def\darr#1{\raise1.5ex\hbox{$\leftrightarrow$}\mkern-16.5mu #1}

\def\half{{\textstyle{1\over2}}} 
\def\roughly#1{\raise.3ex\hbox{$#1$\kern-.75em\lower1ex\hbox{$\sim$}}}



\def\IB{\relax\hbox{$\inbar\kern-.3em{\rm B}$}}
\def\IC{\relax\hbox{$\inbar\kern-.3em{\rm C}$}}
\def\ID{\relax\hbox{$\inbar\kern-.3em{\rm D}$}}
\def\IE{\relax\hbox{$\inbar\kern-.3em{\rm E}$}}
\def\IF{\relax\hbox{$\inbar\kern-.3em{\rm F}$}}
\def\IG{\relax\hbox{$\inbar\kern-.3em{\rm G}$}}
\def\IGa{\relax\hbox{${\rm I}\kern-.18em\Gamma$}}
\def\IH{\relax{\rm I\kern-.18em H}}
\def\IK{\relax{\rm I\kern-.18em K}}
\def\II{\relax{\rm I\kern-.18em I}}
\def\IL{\relax{\rm I\kern-.18em L}}
\def\IP{\relax{\rm I\kern-.18em P}}
\def\IR{\relax{\rm I\kern-.18em R}}
\def\IZ{\relax\ifmmode\mathchoice {\hbox{\cmss Z\kern-.4em Z}}{\hbox{\cmss
Z\kern-.4em Z}} {\lower.9pt\hbox{\cmsss Z\kern-.4em Z}}
{\lower1.2pt\hbox{\cmsss Z\kern-.4em Z}}\else{\cmss Z\kern-.4em Z}\fi}

\def\IB{\relax{\rm I\kern-.18em B}}
\def\IC{{\relax\hbox{$\inbar\kern-.3em{\rm C}$}}}
\def\ID{\relax{\rm I\kern-.18em D}}
\def\IE{\relax{\rm I\kern-.18em E}}
\def\IF{\relax{\rm I\kern-.18em F}}


\def\p{\partial}





\def\half {{1\over 2}}

\def\Lgh{{\varpi_{{}_{\rm{L}}}}}


\def\a{\alpha}
\def\b{\beta}
\def\g{\gamma}  
\def\d{\delta}  
\def\m{\mu}
\def\n{\nu}

\def\l{\lambda} 
\def\k{\kappa}
\def\e{\epsilon}

\def\|{\Big|}
\def\({\Big(}   \def\){\Big)}
\def\[{\Big[}   \def\]{\Big]}







\def\unlockat{\catcode`\@=11}
\def\lockat{\catcode`\@=12}

\unlockat


\def\newsec#1{\global\advance\secno by1\message{(\the\secno. #1)}
\global\subsecno=0\global\subsubsecno=0\eqnres@t\noindent {\bf\the\secno. #1}
\writetoca{{\secsym} {#1}}\par\nobreak\medskip\nobreak}
\global\newcount\subsecno \global\subsecno=0
\def\subsec#1{\global\advance\subsecno by1\message{(\secsym\the\subsecno.
#1)}
\ifnum\lastpenalty>9000\else\bigbreak\fi\global\subsubsecno=0
\noindent{\it\secsym\the\subsecno. #1}
\writetoca{\string\quad {\secsym\the\subsecno.} {#1}}
\par\nobreak\medskip\nobreak}
\global\newcount\subsubsecno \global\subsubsecno=0
\def\subsubsec#1{\global\advance\subsubsecno by1
\message{(\secsym\the\subsecno.\the\subsubsecno. #1)}
\ifnum\lastpenalty>9000\else\bigbreak\fi
\noindent\quad{\secsym\the\subsecno.\the\subsubsecno.}{#1}
\writetoca{\string\qquad{\secsym\the\subsecno.\the\subsubsecno.}{#1}}
\par\nobreak\medskip\nobreak}

\def\subsubseclab#1{\DefWarn#1\xdef #1{\noexpand\hyperref{}{subsubsection}%
{\secsym\the\subsecno.\the\subsubsecno}%
{\secsym\the\subsecno.\the\subsubsecno}}%
\writedef{#1\leftbracket#1}\wrlabeL{#1=#1}}
\lockat

\def\dbend{\lower3.5pt\hbox{\manual\char127}}


\def\boxit#1{\vbox{\hrule\hbox{\vrule\kern8pt
\vbox{\hbox{\kern8pt}\hbox{\vbox{#1}}\hbox{\kern8pt}}
\kern8pt\vrule}\hrule}}

\def\mathboxit#1{\vbox{\hrule\hbox{\vrule\kern8pt\vbox{\kern8pt
\hbox{$\displaystyle #1$}\kern8pt}\kern8pt\vrule}\hrule}}

\overfullrule=0pt


\def\p{\partial}




\def\half {{1\over 2}}

\def\Lgh{{\varpi_{{}_{\rm{L}}}}}


\def\a{\alpha}
\def\b{\beta}
\def\g{\gamma}  
\def\d{\delta}  
\def\m{\mu}
\def\n{\nu}

\def\l{\lambda} 
\def\k{\kappa}
\def\e{\epsilon}

\def\a{\alpha}
\def\b{\beta}
\def\d{\delta}

\def\m{\mu}
\def\n{\nu}

\def\l{\lambda}

\def\k{\kappa}

\def\t{\theta}


\def\|{\Big|}
\def\({\Big(}   \def\){\Big)}
\def\[{\Big[}   \def\]{\Big]}







\def\newdate{9/3/2002} 
 
\def\a{\alpha} 
\def\b{\beta} 
\def\g{\gamma} 
\def\l{\lambda} 
\def\d{\delta} 
\def\e{\epsilon} 
\def\t{\theta}

\def\p{\partial} 
\def\half{{1\over 2}}

  
\Title{\vbox{\hbox{YITP-SB-02-29}}}   
{\vbox{   
\centerline{The Covariant Quantum Superstring}  
\vskip .2cm   
\centerline{and Superparticle from their Classical Actions}}}   
\medskip\centerline
{
P. A. Grassi$^{~a,}$\foot{pgrassi@insti.physics.sunysb.edu},  
G. Policastro$^{~b,c,}$\foot{g.policastro@sns.it},   
and   
P. van Nieuwenhuizen$^{~a,}$\foot{vannieu@insti.physics.sunysb.edu}  
}   
\medskip   
\centerline{$^{(a)}$ {\it C.N. Yang Institute for Theoretical Physics,} }  
\centerline{\it State University of New York at Stony Brook,   
NY 11794-3840, USA}  
\vskip .3cm  
\centerline{$^{(b)}$ {\it Scuola Normale Superiore,} }  
\centerline{\it Piazza dei Cavalieri 7, Pisa, 56126, Italy}  
 \vskip .3cm 
\centerline{$^{(c)}$ {\it New York University, Dip. of Physics,} }  
\centerline{\it 4 Washington Place, New York, NY 10003, USA}  
\medskip  
\vskip  .5cm  
\noindent  
We develop an approach based on the Noether method to construct nilpotent BRST  
charges and BRST-invariant actions. We apply this approach first to the holomorphic part  
of the flat-space  
covariant superstring, and we find that the ghosts $b, c_z$ 
which we introduced by hand in our earlier work, are needed to fix gauge  
symmetries of the ghost action. Then we apply this technique to the superparticle and  
determine its cohomology. Finally, we extend our results to the
combined left- and right-moving sectors of the superstring.  
 
\Date{\newdate}  
  
  
\lref\lrp{  
U.~Lindstr\"om, M.~Ro\v cek, and P.~van Nieuwenhuizen, in preparation.   
}  
  
\lref\pr{  
P. van Nieuwenhuizen, in {\it Supergravity `81},   
Proceedings First School on Supergravity, Cambridge University Press, 1982, page 165.  
}  
  
\lref\polc{  
J.~Polchinski,  
{\it String Theory. Vol. 1: An Introduction To The Bosonic String,}  
{\it String Theory. Vol. 2: Superstring Theory And Beyond,}  
{\it  Cambridge, UK: Univ. Pr. (1998) 531 p}.  
}  
  
\lref\superstring{  
M.~B.~Green and  J.~H.~Schwarz, {\it Covariant Description Of Superstrings,}   
Phys.\  Lett.\ {\bf B136} (1984) 367; M.~B.~Green and J.~H.~Schwarz,  
  {\it Properties Of The Covariant Formulation Of Superstring Theories,}  
  Nucl.\ Phys.\ {\bf B243} (1984) 285\semi  
M.~B.~Green and C.~M.~Hull, QMC/PH/89-7  
{\it Presented at Texas A and M Mtg. on String Theory, College  
  Station, TX, Mar 13-18, 1989}\semi  
R.~Kallosh and M.~Rakhmanov, Phys.\ Lett.\  {\bf B209} (1988) 233\semi  
U. ~Lindstr\"om, M.~Ro\v cek, W.~Siegel,   
P.~van Nieuwenhuizen and A.~E.~van de Ven, Phys. Lett. {\bf B224} (1989)   
285, Phys. Lett. {\bf B227}(1989) 87, and Phys. Lett. {\bf B228}(1989) 53;   
S.~J.~Gates, M.~T.~Grisaru, U.~Lindstr\"om, M.~Ro\v cek, W.~Siegel,   
P.~van Nieuwenhuizen and A.~E.~van de Ven,  
{\it Lorentz Covariant Quantization Of The Heterotic Superstring,}  
Phys.\ Lett.\  {\bf B225} (1989) 44;   
A.~Mikovic, M.~Rocek, W.~Siegel, P.~van Nieuwenhuizen, J.~Yamron and  
A.~E.~van de Ven, Phys.\ Lett.\  {\bf B235} (1990) 106;   
U.~Lindstr\"om, M.~Ro\v cek, W.~Siegel, P.~van Nieuwenhuizen and  
A.~E.~van de Ven,   
{\it Construction Of The Covariantly Quantized Heterotic Superstring,}  
Nucl.\ Phys.\  {\bf B330} (1990) 19 \semi  
F. Bastianelli, G. W. Delius and E. Laenen, Phys. \ Lett. \ {\bf  
  B229}, 223 (1989)\semi  
R.~Kallosh, Nucl.\ Phys.\ Proc.\ Suppl.\  {\bf 18B}  
  (1990) 180 \semi  
M.~B.~Green and C.~M.~Hull, Mod.\ Phys.\ Lett.\  {\bf A5} (1990) 1399\semi   
M.~B.~Green and C.~M.~Hull, Nucl.\ Phys.\  {\bf B344} (1990) 115\semi  
F.~Essler, E.~Laenen, W.~Siegel and J.~P.~Yamron, Phys.\ Lett.\  {\bf B254} (1991) 411\semi   
  F.~Essler, M.~Hatsuda, E.~Laenen, W.~Siegel, J.~P.~Yamron, T.~Kimura  
  and A.~Mikovic,   
  Nucl.\ Phys.\  {\bf B364} (1991) 67\semi   
J.~L.~Vazquez-Bello,  
  Int.\ J.\ Mod.\ Phys.\  {\bf A7} (1992) 4583\semi  
E. Bergshoeff, R. Kallosh and A. Van Proeyen, ``Superparticle  
  actions and gauge fixings'', Class.\ Quant.\ Grav {\bf 9}   
  (1992) 321\semi  
C.~M.~Hull and J.~Vazquez-Bello, Nucl.\ Phys.\  {\bf B416}, (1994) 173 [hep-th/9308022]\semi  
P.~A.~Grassi, G.~Policastro and M.~Porrati,  
{\it Covariant quantization of the Brink-Schwarz superparticle,}  
Nucl.\ Phys.\ B {\bf 606}, 380 (2001)  
[arXiv:hep-th/0009239].  
}  
  
\lref\bv{  
N. Berkovits and C. Vafa,  
{\it $N=4$ Topological Strings}, Nucl. Phys. B {\bf 433} (1995) 123,   
hep-th/9407190.}  
  
\lref\fourreview{N. Berkovits,  {\it Covariant Quantization Of  
The Green-Schwarz Superstring In A Calabi-Yau Background,}  
Nucl. Phys. {\bf B431} (1994) 258, ``A New Description Of The Superstring,''  
Jorge Swieca Summer School 1995, p. 490, hep-th/9604123.}  
  
\lref\OoguriPS{  
H.~Ooguri, J.~Rahmfeld, H.~Robins and J.~Tannenhauser,  
{\it Holography in superspace,}  
JHEP {\bf 0007}, 045 (2000)  
[arXiv:hep-th/0007104].  
}  
  
\lref\bvw{  
N.~Berkovits, C.~Vafa and E.~Witten,  
{\it Conformal field theory of AdS background with Ramond-Ramond flux,}  
JHEP {\bf 9903}, 018 (1999)  
[arXiv:hep-th/9902098].  
}  
\lref\wittwi{  
E.~Witten,  
{\it An Interpretation Of Classical Yang-Mills Theory,}  
Phys.\ Lett.\ B {\bf 77}, 394 (1978);   
E.~Witten,  
{\it Twistor - Like Transform In Ten-Dimensions,}  
Nucl.\ Phys.\ B {\bf 266}, 245 (1986)}  
  
\lref\SYM{  
W.~Siegel,  
{\it Superfields In Higher Dimensional Space-Time,}  
Phys.\ Lett.\ B {\bf 80}, 220 (1979)\semi  
B.~E.~Nilsson,  
{\it Pure Spinors As Auxiliary Fields In The Ten-Dimensional   
Supersymmetric Yang-Mills Theory,}  
Class.\ Quant.\ Grav.\  {\bf 3}, L41 (1986);   
B.~E.~Nilsson,  
{\it Off-Shell Fields For The Ten-Dimensional Supersymmetric   
Yang-Mills Theory,} GOTEBORG-81-6\semi  
S.~J.~Gates and S.~Vashakidze,  
{\it On D = 10, N=1 Supersymmetry, Superspace Geometry And Superstring Effects,}  
Nucl.\ Phys.\ B {\bf 291}, 172 (1987)\semi  
M.~Cederwall, B.~E.~Nilsson and D.~Tsimpis,  
{\it The structure of maximally supersymmetric Yang-Mills theory:    
Constraining higher-order corrections,}  
JHEP {\bf 0106}, 034 (2001)  
[arXiv:hep-th/0102009];   
M.~Cederwall, B.~E.~Nilsson and D.~Tsimpis,  
{\it D = 10 superYang-Mills at O(alpha**2),}  
JHEP {\bf 0107}, 042 (2001)  
[arXiv:hep-th/0104236].  
}  
\lref\har{  
J.~P.~Harnad and S.~Shnider,  
{\it Constraints And Field Equations For Ten-Dimensional   
Super-Yang-Mills Theory,}  
Commun.\ Math.\ Phys.\  {\bf 106}, 183 (1986).  
}  
\lref\wie{  
P.~B.~Wiegmann,  
{\it Multivalued Functionals And Geometrical Approach   
For Quantization Of Relativistic Particles And Strings,}   
Nucl.\ Phys.\ B {\bf 323}, 311 (1989).  
}  
\lref\purespinors{\'E. Cartan, {\it Lecons sur la th\'eorie des spineurs},   
Hermann, Paris (1937)\semi  
C. Chevalley, {\it The algebraic theory of Spinors},   
Columbia Univ. Press., New York\semi  
 R. Penrose and W. Rindler,   
{\it Spinors and Space-Time}, Cambridge Univ. Press, Cambridge (1984)   
\semi  
P. Budinich and A. Trautman, {\it The spinorial chessboard}, Springer,   
New York (1989).  
}  
\lref\coset{  
P.~Furlan and R.~Raczka,  
{\it Nonlinear Spinor Representations,}  
J.\ Math.\ Phys.\  {\bf 26}, 3021 (1985)\semi  
A.~S.~Galperin, P.~S.~Howe and K.~S.~Stelle,  
{\it The Superparticle and the Lorentz group,}  
Nucl.\ Phys.\ B {\bf 368}, 248 (1992)  
[arXiv:hep-th/9201020].  
}  
  
%
%
\lref\GS{M.B. Green, J.H. Schwarz, and E. Witten, {\it Superstring Theory,}   
 vol. 1, chapter 5 (Cambridge U. Press, 1987).    
}  
\lref\carlip{S. Carlip,   
{\it Heterotic String Path Integrals with the Green-Schwarz   
Covariant Action}, Nucl. Phys. B {\bf 284} (1987) 365 \semi R. Kallosh,   
{\it Quantization of Green-Schwarz Superstring}, Phys. Lett. B {\bf 195} (1987) 369.}   
 \lref\john{G. Gilbert and   
D. Johnston, {\it Equivalence of the Kallosh and Carlip Quantizations   
of the Green-Schwarz Action for the Heterotic String}, Phys. Lett. B
{\bf 205}   
(1988) 273.}   
\lref\csm{W. Siegel, {\it Classical Superstring Mechanics},
Nucl. Phys. B {\bf 263} (1986)   
93\semi   
W.~Siegel, {\it Randomizing the Superstring}, Phys. Rev. D {\bf 50} (1994), 2799.  
}     
\lref\sok{E. Sokatchev, {\it   
Harmonic Superparticle}, Class. Quant. Grav. 4 (1987) 237\semi   
E.R. Nissimov and S.J. Pacheva, {\it Manifestly Super-Poincar\'e   
Covariant Quantization of the Green-Schwarz Superstring},   
Phys. Lett. B {\bf 202} (1988) 325\semi   
R. Kallosh and M. Rakhmanov, {\it Covariant Quantization of the   
Green-Schwarz Superstring}, Phys. Lett. B {\bf 209} (1988) 233.}    
\lref\many{S.J. Gates Jr, M.T. Grisaru,   
U. Lindstrom, M. Rocek, W. Siegel, P. van Nieuwenhuizen and   
A.E. van de Ven, {\it Lorentz-Covariant Quantization of the Heterotic   
Superstring}, Phys. Lett. B {\bf 225} (1989) 44\semi   
R.E. Kallosh, {\it Covariant Quantization of Type IIA,B   
Green-Schwarz Superstring}, Phys. Lett. B {\bf 225} (1989) 49\semi   
M.B. Green and C.M. Hull, {\it Covariant Quantum Mechanics of the   
Superstring}, Phys. Lett. B {\bf 225} (1989) 57.}    
 \lref\fms{D. Friedan, E. Martinec and S. Shenker,   
{\it Conformal Invariance, Supersymmetry and String Theory},   
Nucl. Phys. B {\bf 271} (1986) 93.}  
\lref\kawai{  
T.~Kawai,  
{\it Remarks On A Class Of BRST Operators,}  
Phys.\ Lett.\ B {\bf 168}, 355 (1986).}  
 \lref\ufive{N. Berkovits, {\it   
Quantization of the Superstring with Manifest U(5) Super-Poincar\'e   
Invariance}, Phys. Lett. B {\bf 457} (1999) 94, hep-th/9902099.}    
\lref\BerkovitsRB{ N.~Berkovits,   
{\it Covariant quantization of the superparticle   
using pure spinors,} [hep-th/0105050].    
}

  
\lref\BerkovitsFE{  
N.~Berkovits,  
{\it Super-Poincar\'e covariant quantization of the superstring,}  
JHEP { 0004}, 018 (2000)  
[hep-th/0001035].  
}  
  
\lref\BerkovitsPH{  
N.~Berkovits and B.~C.~Vallilo,  
{\it Consistency of super-Poincar\'e covariant superstring tree amplitudes,}  
JHEP { 0007}, 015 (2000)  
[hep-th/0004171].  
}  
  
\lref\BerkovitsNN{  
N.~Berkovits,  
{\it Cohomology in the pure spinor formalism for the superstring,}  
JHEP { 0009}, 046 (2000)  
[hep-th/0006003].  
}  
  
\lref\BerkovitsWM{  
N.~Berkovits,  
{\it Covariant quantization of the superstring,}  
Int.\ J.\ Mod.\ Phys.\ A {\bf 16}, 801 (2001)  
[hep-th/0008145].  
}  
  
\lref\BerkovitsYR{  
N.~Berkovits and O.~Chandia,  
{\it Superstring vertex operators in an AdS(5) x S(5) background,}  
Nucl.\ Phys.\ B {\bf 596}, 185 (2001)  
[hep-th/0009168].  
}  
\lref\BerkovitsZY{  
N.~Berkovits,  
{\it The Ten-dimensional Green-Schwarz   
superstring is a twisted Neveu-Schwarz-Ramond string,}  
Nucl.\ Phys.\ B {\bf 420}, 332 (1994)  
[arXiv:hep-th/9308129].  
}  
  
\lref\BerkovitsUS{  
N.~Berkovits,  
{\it Relating the RNS and pure spinor formalisms for the superstring,}  
hep-th/0104247.  
}  
  
\lref\BerkovitsMX{  
N.~Berkovits and O.~Chandia,  
{\it Lorentz invariance of the pure spinor BRST cohomology   
for the  superstring,}  
hep-th/0105149.  
}  
  
\lref\GrassiUG{  
P.~A.~Grassi, G.~Policastro, M.~Porrati and P.~van Nieuwenhuizen,  
{\it Covariant quantization of superstrings without pure spinor constraints},   
[hep-th/0112162].  
}  
  
\lref\GrassiSP{  
P.~A.~Grassi, G.~Policastro, and P.~van Nieuwenhuizen,  
{\it The massles spectrum of covariant superstrings},   
[hep-th/0202123].  
}  
  
\lref\GrassiWW{  
P.~A.~Grassi, G.~Policastro, and P.~van Nieuwenhuizen,  
{\it Equivalence of the BRST cohomology with/without pure spinors},   
[hep-th/0206216].  
}  
  
\lref\GrassiZZ{  
P.~A.~Grassi, A. Iglesias, G.~Policastro, and P.~van Nieuwenhuizen,  
{\it The Covariant Quantum Superstring and Superparticle from their calssical  
action},   
hep-th/020.  
}

\lref\WittenZZ{  
E.~Witten,  
{\it Mirror manifolds and topological field theory,}  
hep-th/9112056.  
}  
  
\lref\wichen{  
E.~Witten,  
{\it Chern-Simons gauge theory as a string theory,}  
arXiv:hep-th/9207094.  
}  
  
  
\lref\howe{P.S. Howe, {\it Pure Spinor Lines in Superspace and   
Ten-Dimensional Supersymmetric Theories},   
Phys. Lett. B258 (1991) 141, Addendum-ibid.B259 (1991) 511\semi   
P.S. Howe, {\it Pure Spinors, Function Superspaces and Supergravity   
Theories in Ten Dimensions and Eleven Dimensions}, Phys. Lett. B273 (1991)   
90.}  
  
\lref\Oda{I. Oda and M. Tonin, {\it On the Berkovits covariant quantization   
of the GS superstring},   
Phys. Lett. B {\bf 520} (2001) 398 [hep-th/0109051].    
}  
  
\lref\Membrane{  
N. Berkovits,  
{\it Towards covariant quantization of the supermembrane,}  
[hep-th/0201151].  
}  
  
\baselineskip14pt  
  
\newsec{Introduction and Summary}   
  
Recently, a new approach to the completely super-Poincar\'e covariant  
quantization of the superstring with spacetime supersymmetry was  
developed in \GrassiUG \GrassiSP \GrassiWW, based on earlier work by  
Berkovits \BerkovitsFE\ \BerkovitsPH\ \BerkovitsNN\ \BerkovitsWM. A  
free quantum action invariant under BRST transformations and a  
nilpotent BRST generator Q were constructed \GrassiUG. The correct  
massless and massive spectrum for the open and closed string 
was obtained \GrassiSP. The definition of physical states in terms of  
equivariant cohomology was established \GrassiWW.  
In \GrassiUG\ a ghost pair $(c_z,b)$ was introduced  
by hand to make the BRST charge nilpotent, and another BRST-inert  
ghost system (namely $ \eta^m, \omega^m_z$ in \GrassiUG , replaced by  
$\eta^m_z,\omega^m$ in \GrassiSP ) was introduced by hand to  
cancel the central charge. In this article we shall construct the  
quantum action and the BRST charge using the Noether method, and we
obtain in this way a derivation of the ghost pair $b,c_z$. 

We start from the classical Green-Schwarz action, but we take a flat
worldsheet metric\foot{At the tree level the choice of a flat
worldsheet metric is sufficient, but clearly at one loop or for higher
genus surfaces (with or without punctures) it is inadequate.}, and we
replace the $\kappa$ transformation 
$\delta_\kappa \t^\a =\gamma^{\a\b}_m \Pi^m_z \kappa^z_\b$ by the more 
general expression 
$\delta_\lambda \t^\a = \lambda^\a$ where $\lambda^\a$ is a real commuting
$16$-component $D=(9,1)$ spinor.  Using the Noether method applied to
BRST symmetry, new ghosts are added to the action. A preliminary ghost
action will turn out to have a rigid symmetry but is not BRST
invariant. Making this symmetry local leads to the ghost system
$b,\ c_z$ leads and a BRST invariant action.  We apply this general
method to several cases: {\it i)} the heterotic superstring, {\it
ii)} the superparticle and, {\it iii)} the flat space superstring with
combined left- and right-moving sectors. In all the cases we do arrive
at an invariant action and a nilpotent BRST charge.

There exists now a derivation of the $b,c_z$ system from first
principles. For the $\eta_z^m, \omega^m$ ghost system a similar derivaion
is still lacking. 

A different approach, starting from a twisted version of the 
complexified $N=2$ superembedding formulation of the superstring, has been 
studied in 
\lref\MatoneFT{
M.~Matone, L.~Mazzucato, I.~Oda, D.~Sorokin and M.~Tonin,
{\it The superembedding origin of the Berkovits pure spinor covariant  
quantization of superstrings,}
Nucl.\ Phys.\ B {\bf 639}, 182 (2002)
[hep-th/0206104].
}
\MatoneFT. 
  
\newsec{Heterotic Superstring and Superparticle}  
 
The basis for our work is a remarkable identity between the free classical   
({\it i.e.}, without ghosts) superstring $S^{class}_{free}$, the full   
nonlinear classical Green-Schwarz (GS) superstring $S_{GS}$, and antihermitian   
composite objects $d_{L\a}$ and $d_{R \a}$ \csm.     
In the conformal gauge, $h^{\m\n}=\eta^{\m\n}$, one has in Minkowski
space
\eqn\ACTfree{S^{class}_{free} = S_{GS} - \int d^2z   
\Big( d_{L \m\a} (\eta^{\m\n} - \e^{\m\n}) \p_\n \t^\a_L + 
 d_{R \m\a} (\eta^{\m\n} + \e^{\m\n}) \p_\n \t^\a_R \Big) 
}
$$ 
{\cal L}^{class}_{free} = - {1\over 2} \p_\m x^m \p^\m x_m - 
p_{L \m\a} P^{\m\n} \p_\n \t^\a_L -  
p_{R \m\a} \overline{P}^{\m\n} \p_\n \t^\a_R 
$$ 
where 
$P^{\m\n} = (\eta^{\m\n} - \e^{\m\n})$ and $\bar P^{\m\n} = ( \eta^{\m\n} + \e^{\m\n})$.  
Furthermore $S_{GS}=S_{kin}+S_{WZ}$ with  
$$ 
{\cal L}_{kin} = - \half \Pi^m_\m \Pi^\m_m 
$$
\eqn\GSact{
{\cal L}_{WZ}  = - \e^{\mu\nu}\Big[i\, \p_\m x^m\Big(\t_L\g_m\p_\n\t_L-\t_R\g_m\p_\n\t_R\Big)-  
\Big(\t_L\g^m\p_\m\t_L\Big)\Big(\t_R\g_m\p_\n\t_R\Big)\Big] \,  
}
and  
$$d_{L\mu\a } = p_{L\mu\a} + (i \p_\mu x^m + {1\over 2} \t_L
\g^m \p_\m \t_L + {1\over 2} \t_R \g^m \p_\m \t_R) (\g_m \t_L)_\a\,,
$$  
$$
d_{R\mu\a } = p_{R\mu\a} + (i \p_\mu x^m + {1\over 2} \t_L
\g^m \p_\m \t_L + {1\over 2} \t_R \g^m \p_\m \t_R) (\g_m \t_R)_\a\,,
$$
\eqn\pippo{ 
\Pi^m_\m=\p _\m x^m- i\t^\a_L\g^m_{\a\b}\p_\m\t^\b_L - 
i \t^\a_R\g^m_{\a\b}\p_\m \t^\b_R\,.} 
In chiral notation one has ${\cal L}^{class}_{free} = - 1/2 \p x^m
\bar\p x_m - p_{L \a} \bar\p \t^\a_L - p_{R \a} \bar\p\t^\a_R$ with 
$\p = \p_\sigma - \p_t$ and $\bar\p = \p_\sigma + \p_t$. Further, 
$d_{L \a} = p_{L \a} + ( i \p x^m + {1\over 2} \t_L \g^m \p\t_L + 
{1\over 2} \t_R \g^m \p\t_R) (\g_m \t_L)_\a$ and
$d_{R \a} = p_{R \a} + ( i \bar\p x^m + {1\over 2} \t_L \g^m \bar\p\t_L + 
{1\over 2} \t_R \g^m \bar\p\t_R) (\g_m \t_R)_\a \,. $

For us the identity in \ACTfree~is useful becasue it defines objects 
$d_{L \mu \a }$ and $d_{R\mu\a}$ which play a crucial role in what
follows. They become constraints in the quantum theory and form the
starting point for the BRST charge. We denote the left-moving spinor in the  
Green-Schwarz action by  $\t_L$, while  $\t_R$ is the right-moving spinor.  
Chiral $\t$'s have spinorial superscript $\t^\a_L$ and $\t^\a_R$ and  
antichiral $\t$'s are denoted by $\t_{\a}$.  
Thus for the $II A$ case, we use the notation $\t_{\a R}$.  
  
There also exists a relation in Berkovits' approach between the free   
quantum action, the GS action and a BRST exact term. It reads (we  
use the notation $w_\a$ for the conjugate momentum of $\l^\a$ 
instead of $\beta_\a$ of our earlier work to facilitate the  
comparison with \BerkovitsFE\ \BerkovitsPH\ \BerkovitsNN\ \BerkovitsWM) 
\eqn\ACTqu{\eqalign{ 
S^{qu}_{free} & = S_{GS} + Q_B 
\int d^2z   
\Big( w_{L\mu \a} P^{\mu\nu} \p_{\nu} \t^\a_L + w_{R \mu \a}
\bar P^{\m\n} \p_\n \t^\a_R \Big)\,,}} 
where ${\cal L}^{qu}_{free} = {\cal L}^{class}_{free} -
w_{L\mu \a} P^{\mu\nu} \p_{\nu} \l^\a_L - w_{R \mu \a}
\bar P^{\m\n} \p_\n \l^\a_R$. Further $Q_B=(Q_{B,L}+Q_{B,R})$ with  
\eqn\func{
Q_{B,L} = \int d\sigma dt \Big( 
i\l^\a_L {\d \over \d \t^\a_L} + \l_L \g^m \t_L {\d \over \d x^m} + 
d_{L \m} {\d \over \d w_{L\a}} - \Pi^m (\l_L \g_m)_\a {\d \over \d
d_{L\a}} \Big)\,,} and similarly $Q_{B,R}$, which satisfy 
\eqn\bbb{
Q^2_{B,L} = \int d\sigma dt ( -i\l_L \g^m \l_L) 
\Big( {\d \over \d x^m} + (\p_m \t \g^m)_\a 
{\d \over \d d_{L\a}} \Big) - \Pi^m (\l_L \g_m)_\a 
{\d \over \d w_{L\a}} 
} 

In Berkovits approach the BRST operator $Q_B$   
is not hermitian or antihermitean, because his $\l^\a$ is complex, but
in our approach the BRST operator, denoted by Q, is antihermitian.   
For pure spinors $\l$ satisfying $\l\g^m\l=0$, $Q_B$ is clearly
nilpotent on $x^m,\t^\a, \l^\a$ and $d_{z\a}$, but   
does not vanish on $w_\a$. The free quantum action \ACTqu~is invariant under the gauge 
transformation $\d w^\m_\a = \Lambda^\m_m (\g^m \l)_\a$ if 
the $\l$'s are pure spinors, and the BRST operators are nilpotent 
up to a gauge transformation. The $Q_B$ variation of $S_{GS}$ does not vanish   
either, but $S^{qu}_{free}$ is $Q_B$ invariant. The relation in \ACTqu~was discovered   
by Oda and Tonin \Oda, and has been used by Berkovits to construct the pure spinor   
action in a curved background \Membrane. In our derivation below this relation plays no   
role. We shall use the Noether method, applied to BRST symmetry.

In this section we restrict ourselves to one (left-moving) sector (the heterotic
string). In section 4 we discuss the combined left- and right-moving
sector. We start from the GS action which we decompose into a kinetic term and a   
Wess-Zumino (WZ) term, $S_{GS}=S_{kin}+S_{WZ}$. We shall not need $S_{WZ}$   
but only its exterior derivative which is given by the following 3-form both for the II B  
and the II A cases  
\eqn\dWZ{d {\cal L}_{WZ}=-i \, d \t_L\not{\!\!\Pi} d\t_L + i\, d\t_R \not{\!\!\Pi} d\t_R}

The action is invariant under local  $\k$ (Siegel) gauge transformations if one does   
not fix the conformal gauge. We consider the GS action in the conformal gauge.   
In this gauge the $\k$ symmetry transformations acquire extra compensating   
terms and are quite complicated. We  follow therefore a different approach.  
We choose the conformal gauge and replace the composite parameters   
$\not{\!\Pi}\k$ of $\k$ symmetry by a new local classical gauge parameter $\l$.  
The GS action (from now on in the conformal gauge) is of course not invariant   
under the $\l$ transformations of $x^m$ and $\t^\a$, but we shall use the Noether   
method to obtain a BRST invariant free quantum action.  
The new local gauge transformations of $x$ and $\t$ follow straightforwardly  
by replacing $\Pi_{z m} \g^{\a\b} \k^z_\b$  by $\l^\a$ 
\eqn\gauget{\d_\l x^m= -i \l\g^m \t, ~~~\d_\l\t^\a= \l^\a. }  
The matrices $\g^m_{\a\b}$ are real and symmetric, hence the reality 
of $\d_\l x^m$ and of $\d_\l \t^\a$ is preserved. 

The geometrical meaning is at this point unclear. However, \gauget~has the  
same form as the BRST transformations   
generated by the BRST charge $Q_B$ in Berkovits' formalism. Therefore,   
we interpret $\l$ from this point on as a real ghost which changes its statistics:   
$\l$ becomes commuting. The BRST transformations with constant
anticommuting anti-hermitian parameter $\Lambda$ read $\d_B \t^\a = i
\Lambda \l^\a$ and $\d_B x^m = i \Lambda \d_\l x^m$.  
Denoting the BRST transformation of $x^m$ and $\t^\a$ without
$\Lambda$ by $s$, we obtain $s\,\t^\a=i \l^\a$ and $s x^m = \l \g^m \t$. 
The BRST transformations close (they are nilpotent) if  
the $\l$'s are pure spinors. In our approach \GrassiUG~we do not impose any   
constraints on the spinors $\l$, and therefore, to still regain nilpotency of   
the $\l$ transformation, we modify the $\l$ transformation rules of $x$ and   
$\t$ by adding further fields such that they become nilpotent.  
Nilpotency of $s$ is
achieved by defining $s\, \l^\a = 0$, but since $s$ is not nilpotent on
$x$, we introduce a new ghost $\xi_m$ in $s\, x^m$
\eqn\brst{sx^m=\l\g^m\t+\xi^m\,,~~~~~~ s\xi^m=- i\, \l\g^m\l \,,} 
where $\xi_m$ is anticommuting and real. 
We have obtained $s^2=0$ on $x$.   
For the variation of the action we need the variation of $\Pi^m_\m$
which is given by
\eqn\sPi{s\Pi^m_\mu=\p_\m \xi^m+ 2 \l\g^m\p_\m\t \,.}  

The variation of $S_{kin}$ contains a term with a derivative of a ghost   
which we can handle with the Noether approach, and a term with $\p_\m\t$ which  
poses a problem as far as the Noether method is concerned and which therefore  
should be removed   
\eqn\sPiPi{\eqalign{s \Big({1\over 2} \Pi^m_\m\Pi_{\n
m}\Big) = \Pi^m_{(\m}\Big(\p_{\n)}\xi_m+ 2 \l\g_m\p_{\n)}\t\Big)}}  
To remove the term with $\p_\n\t$ we modify the induced metric  
$G_{\m\n} = \Pi^m_{(\mu} \Pi_{\n)m}$ by adding a suitable term to it   
\eqn\Gmunu{\eqalign{G^{mod}_{\m\n}=\Pi^m_{(\m}\Pi_{\n)m}+  2
d_{(\m\a}\p_{\n)}\t^\a.}}  
where $d_{\m\a}$ is a new antihermitian anticommuting field. The extra
term $- d_{\m\a} P^{\mu\nu} \p_\nu \t^\a$ in the action should be
interpreted as a gauge fixing term which breaks the $\kappa$-symmetry. 
The gauge fixed kinetic term varies as follows
\eqn\sG{\eqalign{sG^{mod}_{\m\n}=2\Pi^m_{(\m}\p_{\n)}\xi_m+  
\Big[4 \Big(\l\not{\!\!\Pi_{(\m}}\Big)_\a+ 2 \, sd_{(\m\a}
\Big]\p_{\n)}\t^\a - 2 i\,   
d_{\m\a}\p_\n\l^\a.}}   
  
The most general expression for $sd_{\m\a}$ which leaves only terms with   
derivatives of ghosts is given by  
  
\eqn\sd{\eqalign{s d_{\m\a}=- 2 \Big(\not{\!\!\Pi}_\m\l\Big)_\a+
\p_\m\chi + A_m \Big(\g^m\p_\m\t\Big)_\a}}  
where $A_m$ is an antihermitian
anticommuting vector to be fixed. We used that $\p_{(\mu} \g^m
\p_{\nu)} \t$ vanishes, made a Fierz rearangement and introduced a new
real commuting ghost field $\chi_\a$, which can be interpreted as the
anti-chiral counterpart of the chiral $\l^\a$. 
We fix these free objects by requiring that $sd_{\m\a}$ be $s$ inert   
(nilpotency of $s$ on $d_{\m\a}$).  
This yields   
\eqn\sdschi{\eqalign{sd_\m=\p_\m\chi-2 \not{\!\!\Pi}_\mu\l - 2 i \, \xi^m\g_m\p_\m\t, ~~~~   
s\chi=2 \xi^m\g_m\l.}}  
So far we have achieved that the $s$ variation of  
  \eqn\Lkin{\eqalign{{\cal L}^{mod}_{kin}=-\half\Pi^\m_m\Pi^m_\m -  d_{\m\a}\p^\m\t^\a }}  
contains only terms with derivatives of the ghosts $\l^\a$, $\chi_\a$, and   
$\xi^m$, namely   
\eqn\sLkin{\eqalign{s{\cal L}^{mod}_{kin}=-\Pi^\m_m\p_\m\xi^m- \p^\m\t\p_\m\chi+  
i d_{\m\a}\p^\m\l^\a.}}  
  
We now repeat this program for the WZ term. It is a good consistency check that  
this is possible at all. We define a modified WZ term as follows  
\eqn\LWZ{\eqalign{{\cal L}^{mod}_{WZ}={\cal L}_{WZ} + \e^{\m\n} 
d_{\m\a}\p_\n\t^\a }}  
One finds that also $s{\cal L}^{mod}_{WZ}$ only contains terms with   
derivatives of ghosts  
\eqn\sLWZ{\eqalign{s{\cal L}^{mod}_{WZ}=\e^{\m\n}\Big[
\Pi^m_\m\p_\n\xi_m +   
\p_\m\t\p_\n\chi - i\, d_{\m\a}\p_\n\l^\a\Big].}}  
The sum of all variations is given by 
\eqn\sumofall{
s 
\Big( {\cal L}^{mod}_{kin} + {\cal L}^{mod}_{WZ} \Big) = 
-\Pi^m_\m P^{\m\n} \p_\n \xi_m + i d_{\m\a} P^{\m\n} \p_\n \l^\a -
\p_\mu \t^\a P^{\m\n} \p_\n \chi_\a\,.}  

The next step is to cancel these variations 
by adding free ghost actions and defining suitable   
transformation laws for the antighost fields  
\eqn\Lgh{{\cal L}_{gh}=-\b_{\m m} P^{\m\n}\p_\n \xi^m-w_{\m\a} P^{\m\n}
\p_\n\l^\a- \k^\a_\m P^{\m\n} \p_\n\chi_\a.}  
The antighost $\beta^\m_m$ is anticommuting and anti-hermitian, while
$w_{\m\a}$ and $\k^\a_\m$ are commuting and real. 
Because the variation of ${\cal L}_{kin}+{\cal L}_{WZ}$ contain the operator   
$P^{\m\n}=\eta^{\m\n}-\e^{\m\n}$, the antighosts are holomorphic (chiral on the worldsheet:  
they have the index structure $\b^m_z$,   
$\b_{\a z}$ and $\k^\a_z$). One finds easily a   
particular solution for the variation of the antighosts, 
but the most general solution contains a free
constant $b$ and a target-space bispinor $\eta^{\m,\a\b}$  
  
\eqn\sgh{\eqalign{
s\b^\m_m &=\Big(-\Pi^\m_m- 2 \k^\m\g_m\l\Big)+  
\Big(b\p^\m\xi_m+\half\p^\m b  
\xi_m\Big)+\Big(\chi \eta^\m\g_m\l\Big) \,, \cr 
s w^\m_\a &=   
\Big(i \, d^\m - 2 i \b^\m_m\g^m\l- 2 \xi_m\g^m \k^\m\Big)_\a -i \,   
\Big(b\p^\m\chi_\a+{3\over 4}\p^\m b\chi_\a\Big)+  
\Big(\not\!\xi \eta^\m\chi\Big) \,, \cr 
s \k^{\a\m} &=  
\Big(-\p^\m\t^\a\Big) + i \, \Big(b\p^\m\l^\a+  
{1\over 4}\p^\m b \l^\a\Big)+\big(\eta^\m \not\!\xi \l \Big) \,.}}  
The transformations with $b$ map $\b$ into its own ghost $\xi$ and $w$
and $\kappa$ into the other commuting ghosts 
while the transformations with $\eta^{\m,\a\b}$ map each antighost into the two
non-corresponding ghosts. 

Setting the anticommuting and antihermitian $b$ and the real commuting
$\eta^{\m, \a\b}$ to zero yields a solution of the   
inhomogeneous equation for the transformation laws of the antighosts, but the   
terms with {\it constant} $b$ and  $\eta^{\m,\a\b}$ yield further homogeneous   
solutions. In other words, we are encountering a system with constant   
ghosts-for-ghosts.  
We have already added the terms with a derivative of $b$ for reasons to be   
explained now.   
  
The terms in the transformation rules with constant $b$   
and $\eta^{\m,\a\b}$ yield new rigid symmetries of the ghost action. Although   
we have obtained an $s$-invariant action, the transformation rules for the   
antighosts are not nilpotent. We now let  $b$ become a field  
and add the terms with $\p_\m b$ in \sgh . The action then ceases to be   
invariant, but the transformation laws of the antighosts can be made 
nilpotent by defining suitable transformation laws for $b$ and $\eta$, namely  
\eqn\sbseta{sb=1,~~~~~~~s\eta^{\m, \a\b}=0.}  
In fact the terms in \sgh~with $\eta^{\m,\a\b}$ can be removed by
redefining $\k^{\a\m}  \rightarrow
\k^{\a\m}+ 1/2 (\eta^\m\chi)^\a $ and for this reason we omit them from now
on. This redefinition  leads to a new term in the action of the form
$\chi_\a \eta^{\m,\a\b} \p_\m\chi_\b$; 
however, this extra term is a total derivative which we also omit. 
  
Returning to the problem of making the action BRST invariant, we need   
a kinetic term for $b$. Hence we introduce also a new real
anticommuting ghost $c_\m$ and add the following term to the ghost
action: ${\cal L}^{extra}_{gh} = - b P^{\m\n}\p_\m c_\n$.    
We determine the transformation rule of $c_\m$ such that the
action becomes $s$-invariant. One finds 
\eqn\brstc{
s\, c_\m = - {1\over 2} \Big( \xi^m \p_\m \xi_m - {3 i\over 2} 
\chi_\a \p_\m \l^\a + {i\over 2} \p_\m \chi_\a \l^\a \Big)\,.
}  
Also this transformation law is nilpotent.   

In this way we have reobtained the free BRST invariant action and the nilpotent  
BRST transformation rules of \GrassiUG. In particular we have given a
derivation of the need for the $b, c_\m$ system which follows from the
Noether procedure applied to symmetries of the ghost action. 
However, the problem of giving  a similar fundamental
derivation of the $\eta, \omega$ system remains. For the string the
$\eta, \omega$ system was neeeded to cancel the central charge. For the
superparticle, to which we now turn, the $b,c$ system is needed, but
the $\eta, \omega$ system is not needed because for the superparticle 
there is no central charge and hence we do not need to cancel it. 

\newsec{The superparticle}  

In this section we apply the procedure presented in the previous section  
to the point particle. The operator formalism of \GrassiUG~cannot 
directly be applied in this case becasue $\dot\t$ vanishes
on-shell. The off-shell BRST approach is succesful. 
We consider the open string, hence rigid $N=1$
spacetime susy with one $\t$. We shall show that the correct
spectrum, namely the field equations of $d=(9,1)$ $N=1$ super Yang-Mills theory, is obtained.  
 
We start from the $N=1$ supersymmetric action \lref\brink{L.~Brink and 
J.H.~Schwarz, {\it Quantum Superspace,} Phys. Lett. B {\bf 100}  (1981) 310.} \brink 
\eqn\partA{ 
S = \int d\tau {1\over 2 \, e} \left( \dot x^m - i \t^\a
\g^{m}_{\a\b} \dot \t^\b \right)^2\,, ~~~~ \a = 1, \dots, 16\,, 
} 
which is invariant under $\kappa$-symmetry: 
\eqn\partAA{ 
\d_\k \t^\a = \Pi_m (\g^m \k)^\a\,, ~~~~ \d_\k x^m = {i} \t \g^m \d_\k \t\,,  
~~~~\d_\k e = 4 i \, e\, \dot \t^\a \k_\a\,.  
} 
where $\Pi_m = \dot x^m - {i} \t^\a \g^{m}_{\a\b} \dot \t^\b$.   
The quantization of \partA~is nontrivial because of the fermionic constraint  
$\d S / \d \dot\t^\a = p_\a =  i P^m (\g_m \t)_\a$ with $P_m$ and $p_\a$ the conjugate
momenta to the $x$ and $\t$  
coordinates.  
The anticommutator $\{ p_\a -  {i} P^m (\g_m \t)_\a, p_\b -  {i} P^m (\g_m \t)_\b \}  
= - 2 \g^m_{\a\b} P_m$ shows that the fermionic constraints are both first and second class: only  
half of them anticommute with each other\foot{Decomposing $d_\a =p_\a
- i P^m (\g_m \t)_\a$ into $\not\!P \,d_\a + (1- \not\!P) \, d_\a$, the
$\not\!P\, d_\a$ are first class and the $(1-\not\!P) d_\a$ are second class.}.   
However, it is difficult to disentagle these two classes and construct
a covariant set of independent basis vectors for these 
constraints.\foot{Recently, two of the authors \lref\GrassiQS{ 
P.~A.~Grassi, G.~Policastro and M.~Porrati, 
{\it Covariant quantization of the Brink-Schwarz superparticle,} 
Nucl.\ Phys.\ B {\bf 606}, 380 (2001) 
[hep-th/0009239]. } \GrassiQS presented a solution  
of the quantization of the superparticle using a ``twistor''-like redefinition of variables  
$P^m \g^{\a\b}_m = \l^\a_a (\sigma^+ + P^2 \sigma^-)^a_{~b} \l^{\b b}$ where $\l^\a_a$ are  
the twistor-like variables and $\sigma^{\pm}$ the Pauli matrices. One way to disentagle the two  
types of constraints is an infinite number of ghosts. Using Batalin-Vilkovisky techniques 
the ghosts of level greater than  
three do not interact with the ghost of lower levels and with the other fields of the theory.} 
The theory is invariant under reparametrization of the worldline; however, we will set $e=1$ from  
the beginning and construct a consistent model with local transformation rules.  In the original  
superparticle, one could choose the gauge $e=1$, but then $\kappa$  
transformations acquire extra non-local compensating terms with
$\xi(t) = \int^t dt' (4 i \dot\t k)(t')$.\foot{There  
should be a better way to do this: first go to the light-cone gauge
for the superparticle 
action \partA~and  
reparameterize the fermions by $\zeta^a = \sqrt{p^+} (\g^- \t)^\a$ where 
$\g^\pm = \half (\g^0 \pm \g^9)$. The  
BRST operator for the quantized model is only $Q = c P^2$ and the states are representations  
of the Clifford algebra $\{\zeta^a, \zeta^b\} = 2 \delta^{a,b}$. Berkovits \lref\trieste{N. Berkovits,  
{\it Lectures on Covariant Quantization of Superstrings and Supermembranes},  
ICTP Miramare, Trieste 18-26 March, 2002.} \trieste~finds an interpolating  
BRST operator $\hat Q$ in an enlarged functional space with the unconstrained spinors $\hat\lambda^\a$ and  
their conjugate  momenta $\hat w_\a$, and the composite field $d_\a$. One can show that  
the cohomology can be constructed in two equivalent ways: the first reproduces the light-cone massless  
states of the superparticle, the other reproduces the BRST cohomology with pure spinor constraints. It  
would be interesting to repeat this approach for our formulation. }  
 
We compute the variation of \partA~under the BRST transformations 
\eqn\partB{ 
s\, x^m = \xi^m + \t \g^m \l\,, ~~~~~ s\, \t^\a =i\, \l^\a\,, ~~~~  
s\, \xi^m = - i \l \g^m \l \,, ~~~~s\, \l^\a = 0\,.  
} 
In order that the variation of \partA~be proportional to the equations of motion  
of the ghost fields, we add the term $\int d\tau d_\a \dot\t^\a$ where $d_\a$ and  
its BRST variation are given by  
\eqn\partC{ 
d_\a = p_\a + i \dot x_m (\g^m \t)_\a + {1\over 2} (\g^m \t)_\a (\t \g_m \dot \t)\,,  
} 
$$ 
s\, d_\a = \dot \chi_\a - 2\, \Pi_m \g^m \l + \Lambda_m (\g^m \dot\t)_\a  
$$ 
where $\Lambda_m$ and $\chi_\a$ are two arbitrary fields.  
Notice that we can freely add the ghost $\chi_\a$ since on-shell  
this term vanishes. The BRST transformation  
of $d_\a$ is nilpotent if 
$$\Lambda_m = - 2 i \xi_m\,, ~~~~s\, \chi_\a = 2 \xi^m (\g_m \l)_\a$$  
Then, following the procedure already discussed, we add   ghost terms  to the action
\eqn\partD{ 
S_{gh,1} = \int d\tau \left( \beta_m \dot\xi^m + w_\a \dot\l^\a + \k^\a \dot\chi_\a \right)  
} 
whose variation cancels against the variation of $S + \int d\tau d_\a \dot\t^\a$ if the  
antighosts transform in the following way 
\eqn\partE{ 
s\, \b_m =- \Pi_m - 2 \k \g_m\l + b \dot\xi_m+\half \dot b \xi_m \,,  } 
$$ 
s\, w_\a = i\, d_\a - 2 i\, \b_m (\g^m \l)_\a  - 2 
\xi_m (\g^m \k)_\a - i\,  b \dot \chi_\a - {3 i \over 4}\dot b \chi_\a  \,,   
$$ 
$$ 
s\, \k^{\a}  = - \dot \t^\a + i b \dot\l^\a + {i\over 4} \dot b \l^\a  \,.  
$$ 
The contributions with ghosts-antighosts in the transformation rules  
are needed to compensate the non-linear variations of the
ghost fields $\xi^m$ and $\chi_\a$ in the action \partD.  
Further the terms proportional to $b$ or $\dot b$ are needed  
to obtain a nilpotent BRST symmetry. As we learned from the previous section, a suitable redefinition of  
$\kappa^\a$ removes the $\eta^m$ terms from the symmetry, therefore we have already chosen the basis  
without $\eta^m$. The nilpotency of the BRST symmetry is achieved by defining $s\, b = 1$.  
 
The last step is to add  a $b-c$ term to the action and derive 
the BRST transformation for the ghost $c$  
\eqn\partF{ 
S_{gh,2} = \int d\tau b \dot c\,,~~~~~~~  
s\, c = -\half \left(\xi^m \dot \xi_m - {3i \over2} \chi_\a \dot\l^\a
+ {i\over2} 
\dot\chi_\a \l^\a \right)\,.  
} 
The sum $S + S_{gh,1} + S_{gh,2}$ is now invariant under BRST symmetry.  
At this point, we can rewrite the terms of the action which contain the field $x^m$ in a  
first order formalism. Namely, $\int d\tau \half \Pi^2 =  
\int d\tau (P_m \Pi^m - \half P^2)$. Canonical quantization implies that  
$[P^m, x^n] =- i \eta^{mn}$.  This will be used in the next section. 
 
We now turn to the determination of the massless cohomology for the superparticle.   
The physical states of the superparticle should be found at ghost number $1$.  
Without further restriction, the  
cohomology is however trivial, but following \GrassiSP~we  
assign a grading to the ghost fields 
\eqn\partG{ 
gr(\lambda^\a) = 1\,, ~~~~ gr(\xi^m) =2\,, ~~~~ gr(\chi_\a) =3\,, ~~~~ gr(c) =4\,, 
} 
and the corresponding opposite numbers for antighosts. We cannot use
the affine Lie algebra to determine the grading
of $\chi$ and $c$ as in \GrassiSP, because $\dot \theta =0$ is a here a field equation
and there is no central charge for a point particle. However,
observing that the part $Q_0$ of the BRST operator which only contains
ghost and antighost fields  is nilpotent by itself, one can introduce
a grading which explains this. Namely $Q_0$ has vanishing grading and
this yields $gr(\chi) =3$ and $gr(b)= -4$. The relevant cohomology  
is selected in the functional space of non-negatively graded polynomials denoted in the  
following by ${\cal H}_+$.\foot{Notice that 
in the pure spinor formulation, $\l^\a$ should be complex and its complex conjugate $\bar\l_\a$  
should transform under the conjugated representation of $Spin(9,1)$. This implies that one can construct a homotopy  
operator ${\cal K}$ for the BRST charge $Q_B = \l^\a d_\a$. It is easy to show that  
${\cal K} = \bar\lambda_\a \t^\a / (\bar\l \l)$ with $(\bar\l \l) =  \bar\lambda_\a \l^\a$ satisfies $\{ Q, {\cal K}\} = 1$.  
This obviously renders the cohomology  
in \BerkovitsRB~trivial since every $Q$-closed expression is also $Q$-exact.  
In order to obtain a nontrivial cohomology one may use the grading in \partG~and observe  
that the homotopy operator ${\cal K}$ has negative grading.} 
 
The most general scalar expression in ${\cal H}_+$ with ghost number one is  
\eqn\cohoC{\eqalign{ 
{\cal U}^{(1)}(z) &= i \l^\a A_\a + \xi^m A_m + \chi_\a W^\a  \cr 
& + b\, \Big( \xi^m \xi^n F_{m n}   + i \l^\a \chi_\b F^{~~\b}_\a + \chi_\a \, \xi^m F^{\a}_{~~m}  
+ \chi_\a \chi_\b F^{\a \b} \Big) \,,  
}} 
where $ A_\a, \dots, F^{\a\b}$ are arbitrary superfields depending on $x_m, \t^\a$. The requirement of   
positive grading has ruled out $b \l^\a \l^\b$ and $b \l^\a \xi^m$.  
 
The condition $\{ Q, {\cal U}^{(1)}(z) \} = 0$ implies the following equations  
\eqn\cohoD{\eqalign{  
& D_{(\a} A_{\b)} + i \g^m_{\a\b} A_m = 0 \,, \cr  
& \p_m A_\a - D_\a A_m - 2 i\, \g_{m\a\b} W^\b = 0 \,, \cr  
& \p_{[m} A_{n]} + F_{mn} = 0 \,,  \quad\quad 
D_\b W^\a + F_\b^{~~\a} = 0 \,, \cr 
& \p_m W^\a + F_{~~m}^{\a} = 0 \,, \quad\quad 
F^{\a\b} = 0 \,, \cr 
}} 
where $D_\a \equiv \p / \p \t^\a -i \t^\b \g^m_{\a\b} \p / \p x^m$ 
\foot{
Notice that $D_\a$ is hermitian. We define $D_{(\a} A_{\b)} =  
\half \left( D_\a A_\b + D_\b A_\a \right)$ and $\p_{[m} A_{n]} =  
\half \left( \p_m A_n - \p_n A_m \right)$.}. 
The terms in  $\{ Q, {\cal U}^{(1)}(z) \} $ which contain the field $b$ yield  
equations which are the Bianchi identities \GrassiUG.  
From the first two equations of \cohoD~one gets the field  
equations for $N=1, d=(9,1)$ super-Maxwell theory   
\eqn\neh{\eqalign{ 
& \g_{[mnpqr]}^{\a\b} D_\a A_\b = 0 \,,  
\cr}}  
as well as the definition of the vector potential $A_m$ and  
the spinorial field strength $W^\a$ in terms of $A_\a$ 
\eqn\nehB{\eqalign{ 
A_m = {1\over 16} \g_m^{\a\b} D_\a A_\b \,, \quad \quad 
W^\a = {1\over 20} \g_m^{\a\b} \left( D_\b A_m - \p_m A_\b \right)\,. }    
} 
Moreover, the remaining equations in \cohoD~ imply that the curvatures  
$F_{mn}, F^{\a}_{~~m}$, and $F_{\b}^{~~\a}$ are expressed in terms of the  
spinor potential $A_\a$.  
 
The gauge transformations of the vertex ${\cal U}^{(1)}(z)$ are generated by the BRST  
variation of a spin-zero ghost-number-zero  
field $\Omega^{(0)}(z) \in {\cal H}_+$, whose most  
general expression is given by  $\Omega^{(0)}(z) = C$,  
with  $C $ arbitrary superfield. The BRST variation of   $\Omega^{(0)}$ is   
$\delta {\cal U}^{(1)}(z) = \Big[ Q, \Omega^{(0)}(z) \Big] = i \l^\a
D_\a C + \xi^m \p_m C$.  
One can easily check that $C$ is the usual parameter of the gauge transformations on  
the super-Maxwell potentials: $\delta A_\a = D_\a C, ~~\delta A_m = \p_m C$.  
Thus, the only independent superfield is $A_\a$, and it satisfies \neh~which is gauge invariant.  
For further discussion of these field equations we refer to \GrassiUG.  


\newsec{Closed Superstrings}  
 
In this section we again apply the procedure of section 2, but now to the combined left-moving  
and right-moving sector of the Green-Schwarz superstring simultaneously.   
 
We start from the GS  action in \GSact~. The transformation rules are
now given by 
\eqn\brssymm{\eqalign{  
&s\, x^m = (\t_L \g^m \l_L + \xi_L) + (\t_R \g^m \l_R + \xi_R)\,,
~~~~~~ \cr  
&s\, \t^\a_L = i \l^\a_L\,, ~~~~~~~~ s\, \t^{\hat\a}_R = i \l^{\hat\a}_R\,, \cr  
&s\,  \l^\a_L =  s\, \l^{\hat\a}_R = 0\,,~~~~~~~   \cr
&s\, \xi^m_L = -i \l_L \g^m \l_L\,,~~~~~~~   
s\, \xi^m_R = -i \l_R \g^m \l_R \,, \cr  
}}    
One clearly has nilpotency on these fields. 

Next we add to ${\cal
L}_{GS}$ the terms with $d_{L z\a} \equiv d_{L,1 \a} - d_{L,0 \a}$ and 
$d_{R \bar z\a} \equiv d_{R,1 \a} + d_{R,0 \a}$
\eqn\newDD{
{\cal L}_d = - d_{L z \a} \bar \p \t^\a_L - 
d_{R \bar z \a} \p \t^\a_R\,.
}
We recall that $d_{L z \a}$ and $d_{R \bar z\a}$, given below 
\pippo, are such that in ${\cal L}_{GS} + {\cal L}_d$ only the 
free kinetic terms for $x, \t_{L/R}$ and $p_{L/R}$ remain. As before we
determine the variations of $d_{L z \a}$ and $d_{R \bar z\a}$ (hence
of $p_{L z \a}$ and $p_{R \bar z\a}$) by requiring that 
in the $s$-variation of ${\cal L}_{GS} + {\cal L}_d$ the terms without 
derivatives of ghosts cancel. However, we also require nilpotency on 
$d_{L z \a}$ and $d_{R \bar z\a}$; since there are cross-terms, this
is less trivial. We find it convenient to introduce an
auxiliary field for $\Pi^m_0$, so we replace $ 1/2\, (\Pi^m_0)^2$ by 
$- 1/2 P_0^m P_{0 m} + P^m_0 \Pi_{0 m}$. There are now two ways to
proceed 

{\it i)} we take the rules of the heterotic string in each
sector, but the cross-terms in $s d_{L z \a}$ are determined by requiring nilpotency on $P^m_0$ and
$d_{L z \a}$. One can achieve this, but one has then only nilpotency on
$d_{R z \a}$ modulo the free field equations of $\t_{L/R}$ and
$\xi_{L/R}$. 

{\it ii)} We write all transformation rules with only $\p_1$
derivatives, but not with any $\p_0$ derivatives. This can be achieved
by using the free field equations. This changes the rules of the
heterotic string, but we obtain nilpotency on all fields. 

Since one either works with the heterotic string or with the Green-Schwarz
string, we adopt the second procedure. We obtain then 
\eqn\ddtran{\eqalign{
&s\, d_{L z \a} = 2 \p_1 \chi_{L\a} - 2 (\Pi_{1 m} - P_{0 m}) 
\g^m_{\a\b} \l^\a_L - 4 i \xi_{L m} \g^m_{\a\b} \p_1 \t^\b_L \,, \cr  
&s\, d_{R \bar z \a} = 2 \p_1 \chi_{R\a} - 2 (\Pi_{1 m} + P_{0 m}) 
\g^m_{\a\b} \l^\a_R - 4 i \xi_{R m} \g^m_{\a\b} \p_1 \t^\b_R \,, \cr  
&s\, P^m_0 = - 2 ( \l_L \g^m \p_1 \t_L - \l_R \g^m \p_1\t_R) - \p_1
\xi^m_{L} + \p_1 \xi^m_{R}\,, \cr  
&s\, \Pi^m_1 = 2 \l_L \g^m \p_1 \t_L + 2 \l_R \g^m \p_1\t_R + \p_1
\xi^m_{L} + \p_1 \xi^m_{R}\,, \cr
& s \chi_{L \a} = 2 \xi^m_L (\g_m \l_L)_\a\,, \cr
& s \chi_{R \a} = 2 \xi^m_R (\g_m \l_R)_\a\,.
}}
It is clear that nilpotency of $s$ holds on $\Pi^m_1$, $\Pi^m_0$ and $P^m_0$ in each
sector separately. We have written $s\, \Pi^m_1$ below $s\, P^m_0$ so
that the difference becomes clear: in $s\, P^m_0$ we have used the
field equations $(\p_1 + \p_0)\t^\a_L =0\,,~~
(\p_1 - \p_0)\t^\a_R =0$, 
$(\p_1 + \p_0)\xi^m_L =0\,,~~(\p_1 - \p_0)\xi^m_R =0$. Because there
are only $\p_1$ derivatives in $\Pi^m_1$ and $P^m_0$, 
nilpotency of $s d_{L z \a}$ and $s d_{R \bar z \a}$ is 
relatively easy to prove. 

Using these transformation rules, one finds 
\eqn\varyactA{ 
s\, S = \int d^2z \Big[( P^m_0 - \Pi^m_{1}) \bar\p \xi_{Lm} - (
P^m_0 + \Pi^m_{1}) \p \xi_{Rm}}
$$    
- 2 \p_1  \chi_{L\a}\bar\p \t^\a_L - 2 \p_1  \chi_{R\a} \p \t^\a_R +     
i d_{L z \a} \bar\p \l^\a_L +  i d_{R\bar z \a} \p \l^{\a}_R \Big]\,.  
$$  
To prove this simple result requires multiple partial integrations and
Fierz identities. To cancel these variations we add the ghost action 
\eqn\ghostac
{  
S_{gh,1} = \int d^2z \Big(
w_{L z\a} \bar\p \l^\a_L +  w_{R \bar z\a} \p \l^{\a}_R +   
\b_{L z m} \bar\p \xi^m_L +  \b_{R \bar z m} \p \xi^{m}_R +   
\k^\a_{L z} \bar\p \chi_{L\a} +  \k^{\a}_{R \bar z} \p \chi_{R\a}\Big)
}  
and choose the appropriate transformation laws for the antighosts  
$$  
s\, w_{L \a} = - i d_{L \a} -
 2 i \, \b_{L m} (\g^m \l_L)_\a - 2 \xi_{L m} (\g^m \k_{L})_\a + 2 i
b_L \p_1 \chi_{L\a}  + {3 i \over2} \p_1 b_L \chi_{L\a}\,,   
$$  
\eqn\brstanti{\eqalign{  
s\, \b_{Lm} =  -  P_{0 m} + \Pi_{1 m} - 2 \k_{L} \g^m \l_L -  2 b_L
\p_1 \xi_{Lm} - \p_1 b_L\,  \xi_{Lm}\,,  
}}  
$$  
s\, \k^\a_L = 2 \p_1 \t^\a_L -  2 i  b_L \p_1  \l^\a_L - {i\over 2} 
\p_1 b_L\,  \l^\a_L \,.  
$$  
The rules for the right-moving antighosts $w_{R\a}, \b^m_R$ and
$\kappa^\a_R$ are obtained by replacing $- P^m_0$ by $P^m_0$ (and $L$
by $R$ of course). These rules are nilpotent if $s\, b_L = s\, b_R =
1$, but the action is not yet invariant. Since it varies into term
with $b$ we add the ghost action
\eqn\ghostghost{  
S_{gh,2} = \int d^2z \Big[  b_{L} \bar\p c_L +  b_{R} \p c_R \Big]}  
and find the transformation rules for $c_L$ and $c_R$ from the BRST
invariance of the action
\eqn\transC{  
s\, c_L = - \xi_L \p_1 \xi_L + {3 i \over 2} \chi_{L\a} \p_1 \l^\a_L -
{i\over 2} 
\p_1\chi_{L\a} \l^\a_L 
}  
and, analogously, for $c_R$. Nilpotency only fixes the terms with
$\p_1 b_L$ in \brstanti~up to an overall constant, but invariance of the
action fixes this constant. All transformation rules for the combined
sectors are now nilpotent; this has been achieved by introducing only
one auxiliary field, namely $P^m_0$. 

Needless to say, we can again define the grading   
current and we define the BRST cohomology on the space of   
non-negatively graded vertices.   

  
\newsec{Acknowledgements}  
We thank E. Cremmer for l' hospitalit\'e de 
l' Ecole Normale Sup\'erieure where part of this work has been 
done. This work was partly funded by NSF Grant PHY-0098527. 
  
  
\footatend\vfill\supereject\immediate\closeout\rfile\writestoppt
\baselineskip=14pt\centerline{{\bf References}}\bigskip{\frenchspacing%
\parindent=20pt\escapechar=` \input refs.tmp\vfill\eject}\nonfrenchspacing  
  
\bye